# EFFECT OF EYE DOMINANCE ON THE PERCEPTION OF STEREOSCOPIC 3D VIDEO

*Amin Banitalebi-Dehkordi[1], Student Member, IEEE, Mahsa T. Pourazad[1,2], Member, IEEE, and Panos Nasiopoulos[1], Senior Member, IEEE*

[1]Department of Electrical & Computer Engineering, University of British Columbia, Canada
[2]TELUS Communications Inc., Canada
{dehkordi, pourazad, panosn}@ece.ubc.ca

**ABSTRACT**

Asymmetric schemes have widespread applications in the 3D video transmission pipeline. The significance of eye dominance becomes a concern when designing such schemes. In this paper, in order to investigate the effect of eye dominance on the perceptual 3D video quality, a database of representative asymmetric stereoscopic sequences is prepared and the overall 3D quality of these sequences is evaluated through subjective experiments. Experiment results showed that viewers find an asymmetric video more pleasant when the view with higher quality is projected to their dominant eye. Moreover, the eye dominance changes the mean opinion quality score by 16 % at most, a result caused by slight asymmetric video compression. For all other representative types of asymmetry, the statistical difference is much lower and in some cases even negligible.

***Index Terms—*** Eye dominance, asymmetric stereoscopic video, 3D video quality of experience.

## 1. INTRODUCTION

Three-dimensional video technologies have started penetrating the consumer market in recent years. Meanwhile, one important objective of the broadcasting industry is to deliver 3D content to the end-users in a proper display format. In order to provide end users with a high quality 3D experience, a limited number of views (and the corresponding depth map sequences) of a scene are encoded and transmitted [1]. At the decoder side, additional views are synthesized using the available views (and depth maps) to support autostereoscopic multiview displays. Given the demand of additional bandwidth, several asymmetric schemes have been proposed for transmitting multiview video content. The basic idea here is that although the viewer is presented with stereoscopic content that includes low quality views, the high quality views could mask the low quality of the other views by taking advantage of the binocular suppression theory [2].

Asymmetric schemes achieve high compression performance through allocating low bitrate to a selected number of views while compressing the rest of views at higher bit rate (asymmetric compression). One way of achieving this is by compressing some of the views using a low Quantization Parameter (QP) while a higher QP level is used for compressing the rest of the views. It has been shown that this scheme outperforms the convenient symmetric video coding in terms of bitrate [3-6]. Another way of achieving asymmetric compression is to blur a selected number of views before compression so that their bitrate is reduced. Note that blurring high frequency details results in higher compression [7]. Another asymmetric approach is to reduce the resolution of a selected number of views by down-sampling them before compression. At the decoder side these views are up-sampled to their original resolution [8-9]. In order to support a wide range of multiview displays, view synthesis is performed at the receiver-end to generate the required additional views. As a result, the generated multiview content will include some stereo pairs, which consist of one original view, and one synthesized view (asymmetric content) [10].

While the use of asymmetric schemes for compressing and displaying multiview video content is growing, a major question arises regarding the effect of such schemes on the 3D quality of experience (QoE). As reported in [11], 70% of humans are right-eye dominant, 20% are left eyed, and 10% have no eye preference. Displaying asymmetric 3D content to viewers with different eye dominance is likely to affect binocular depth-cues perception and thus their 3D QoE.

Several subjective studies focused on understanding the significance of eye dominance. Meegan et al. performed a psycho-visual study on the quality of asymmetric stereo images which showed that although blurring has no effect on the overall 3D quality, compression blockiness degrades the perceived quality [12]. Unlike Meegan's study, Seuntiens et al. reported that eye dominance has no impact on the quality of stereo images when the pictures are encoded using asymmetric JPEG compression [13]. In the case of

This work was supported in part by NSERC under Grant STPGP 447339-13 and the ICICS/TELUS People & Planet Friendly Home Initiative at UBC.

stereoscopic videos, Jain et al. reported that according to their experiments blurring does not change the perceived 3D quality [14].

In this paper we investigate the relationship between eye dominance and the 3D quality of experience when 3D videos have asymmetric distortions due to asymmetric compression or the displaying technology used. To this end, we create a database of stereo pairs with representative asymmetric distortions. These distortions include blurring, compression (asymmetric video coding), down-sampling (mixed resolution stereo pair), and synthesizing. Then the effect of eye dominance on the quality of asymmetric 3D videos is studied through a series of subjective tests.

The rest of this paper is organized as follows: Section 2 describes the subjective evaluation process as well as the utilized video dataset, Section 3 discusses the results, and Section 4 concludes the paper.

## 2. ASYMETRIC VIDEO QUALITY EVALUATION

This section elaborates on the preparation of the asymmetric video database and subjective experiments.

### 2.1. Representative asymmetric video dataset

Four original stereoscopic sequences are chosen from the sequences proposed by MPEG for common 3D video quality tests [15] for our study. Table I provides the details on the specifications of these sequences.

To study the viewers' QoE when watching asymmetric content, we need to create representative asymmetric stereo pairs. To this end, the following distortions are applied to the test stereo video sequences:

#### 2.1.1. Asymmetric video compression
Asymmetric video compression is one way to reduce bandwidth. In our implementation we apply compression to only one of the views. In other words, for each stereoscopic video, two sets of simulated asymmetric compressed videos are generated, where only the right or left view is compressed and the other view is uncompressed.

The High Efficiency Video Coding (HEVC) standard (HM version 9) is used, with GOP (group of pictures) size of 8, and the random access high efficiency configuration, ALF, SAO, and RDOQ enabled [16-18]. To ensure the compression distortions are visible enough, several asymmetric compressed stereo videos are created using different Quantization Parameter (QP) values. Subjective evaluations showed that QP levels of 40 and 50 result in noticeable and severe visual artifacts, respectively. These QP levels are the ones used in our evaluations.

#### 2.1.2. Blurring one view
Blurring a picture before encoding it improves the compression performance, as it removes the high frequency details of the picture. In this asymmetric case, the two views are blurred at different levels (severe and negligible). For the lower blurring level, a Gaussian kernel with size of 16×16 and standard deviation of 8 is used, while higher blurring effect is imposed to the other view using a Gaussian kernel with size of 32×32 and standard deviation of 16. For each stereo pair, two sets of simulated asymmetric blurred videos are generated where only right or left view is blurred using these filters and the other view is the original one.

#### 2.1.3. Mixed resolution
In a mixed resolution stereoscopic video, the resolution of one view is lowered down so that it is compressed to a lower bit rate [8-9]. At the decoder side, the low resolution view is up-scaled to its original size. The down-sampling factors of 8 and 32 are used in this study. Empirically it is found that when one of the views is down-sampled by factor of 8 the 3D quality degradation is at the threshold of becoming noticeable while a down-sampling factor of 32 results in poor quality.

#### 2.1.4. View synthesis
The only way to provide compatibility with every type of present and future multiview display technologies is by synthesizing the required number of views at the receiver-side, using the limited number of transmitted views and their corresponding depth maps. In this scenario the generated multiview content includes stereo pairs that consist of one original view and one synthesized one (asymmetric content). In our implementation we synthesize views using the MPEG view synthesis reference software (VSRS) [19].

### 2.2. Subjective quality evaluation

Our asymmetric video database contains 56 generated videos (16 with one view blurred, 16 with asymmetric compression, 16 with mixed resolution, and 8 with one view synthesized). The quality of these videos is subjectively evaluated by two groups of subjects, one consisting of eighteen right-eyed dominant subjects and the other of sixteen left-eyed dominant subjects.

TABLE I
SPECIFICATIONS OF THE 3D VIDEO DATABASE

| Sequence | Resolution | Frame Rate (fps) | Number of Frames | Spatial Complexity | Temporal Complexity | Depth Range | Motion Level | Input Views | View to Synthesize |
|---|---|---|---|---|---|---|---|---|---|
| Poznan_Hall2 | 1920×1080 | 25 | 200 | Medium | Medium | Medium | Medium | 7-6 | 6.5 |
| Poznan_Street | 1920×1080 | 25 | 250 | High | High | High | High | 4-3 | 3.5 |
| GT_Fly | 1920×1080 | 25 | 250 | High | High | High | High | 5-2 | 4 |
| Dancer | 1920×1080 | 25 | 250 | Medium | Medium | Medium | Medium | 2-5 | 3 |

The Single Stimulus (SS) method was used, where quality of each video is rated independently and not in comparison to any reference video. Grading is based on the Numerical Categorical Judgment (NCJ) method [20] which uses 11 discrete rating scores from 0 (lowest score) to 10 (highest quality). A HYUNDAI 46" 3D TV display with passive polarized glasses was used.

After running the experiments and collecting the scores, outlier detection is performed according to the ITU BT.500-13 recommendation [20]. There was no outlier among left-eye dominant subjects; however, there were two right-eye dominant outliers, which were removed from any further data analysis.

## 3. RESULTS AND DISCUSSIONS

Once the results of the subjective experiments were collected, Mean Opinion Score (MOS) was calculated for each video as the average of the scores associated to that video with a 95 % confidence interval. Then, the MOS values were reported for right-eye dominant and left-eye dominant subjects separately. Fig. 1.a shows the mean opinion score reported for left and right eye dominant subjects when only the left view is distorted. Note that by "distortion" or "artifact" we refer to the asymmetry types that are mentioned in Section 2.1. Similarly, the perceptual quality when only the right view is distorted is reported in Fig. 1.b. As it is observed from Fig. 1.a, when the left view is distorted, left eye dominant observers give a lower score to the videos. This means that the distortions are more visible to them since their dominant eye is watching the distorted view. Similarly, when the right view is distorted (Fig. 1.b), right-eye dominant viewers notice the distortions and assign lower quality scores.

The overall 3D quality of experience with 95% confidence interval observed by the left-eye dominant subjects is illustrated in Fig. 2.a; this represents cases where only either the left view or the right view is distorted. As it is observed, there is a slight preference towards the case where only the right view is distorted, as the artifacts in the left view are more visible. Fig. 2.b demonstrates the MOS of the right-eye dominant observers for cases where either only the left view or the right view is distorted. It can be observed that the right-eye dominant subjects show slight preference towards the case where only the left view is distorted (opposite to Fig 2.a).

Results from Fig. 1 and Fig. 2 indicate that the difference between the reported MOS of right-eye dominant and left-eye dominant subjects is negligible when the distortions are severe (i.e., when one view is highly blurred, compressed, or up-sampled from a very low resolution). In these cases, the quality of the distorted view is so poor that regardless of the quality of the other view, the overall quality becomes poor. Therefore, no matter which eye is dominant, the video would appear low quality to the observers. On the contrary, in the case that videos are slightly distorted, i.e., the distortions are approximately at the threshold of being noticed, the difference between the reported quality by left and right eye dominant observers is higher. The reason is that, in this case, distortions are perceived only if they are exposed to the dominant eye.

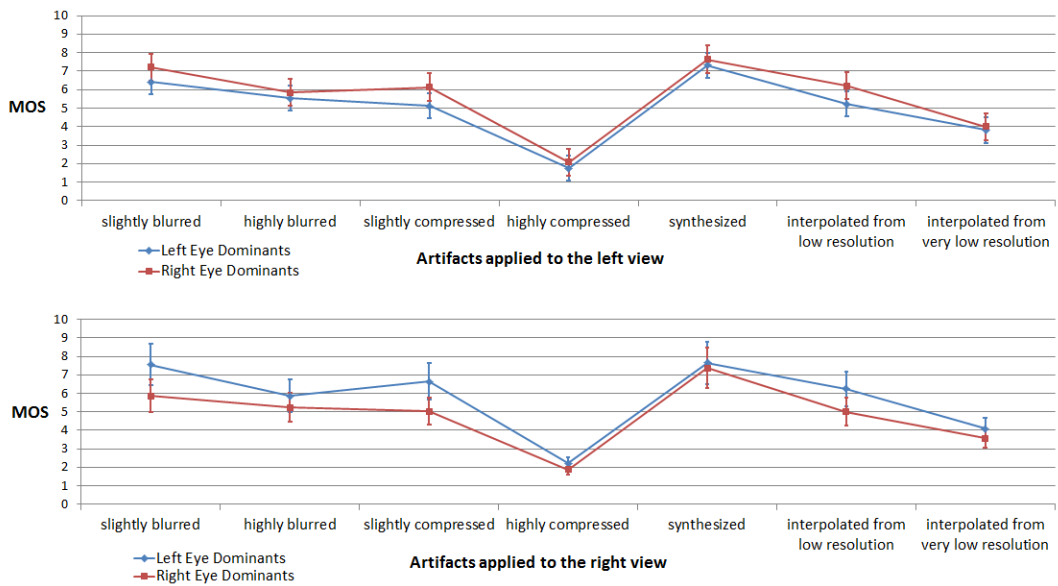

**Fig. 1. Video quality for left/right eyed observers when only left view (a) or right view (b) is distorted.**

TABLE II
STATISTICAL DIFFERENCES BASED ON STUDENT T-TESTS: P-VALUES

| P-Values | Slightly blurred | Highly blurred | Slightly compressed | Highly compressed | Synthesized | Interpolated from low resolution | Interpolated from very low resolution |
|---|---|---|---|---|---|---|---|
| Fig. 1.a | 0.68 | 0.61 | **0.01** | 0.54 | 0.55 | 0.60 | 0.16 |
| Fig. 1.b | 0.11 | 0.21 | **0.07** | 0.15 | 0.34 | 0.75 | 0.18 |
| Fig. 2.a | 0.73 | 0.31 | **0.01** | 0.74 | 0.28 | 0.44 | 0.22 |
| Fig. 2.b | 0.18 | 0.11 | **0.01** | 0.17 | 0.63 | 0.48 | 0.32 |

To investigate if the scores assigned by left and right eye dominant subjects demonstrate significant statistical difference, P-values are calculated for each type of artifact. As observed from Table II, eye dominance affects the 3D quality in the case of slight asymmetric video compression (P-values close to zero). For the rest of the asymmetric cases, however, no statistical difference is observed.

Based on these results, we conclude that, in general, eye dominance can change the overall perceived quality of a stereoscopic video. However, depending on the type and severity of the asymmetry, the effect of eye dominance on the overall perceived quality will change. Specifically, the eye dominance can affect the perceived quality by up to 16 % in the case of asymmetric video compression when one of the views is slightly compressed. One way to compensate for the effect of eye dominance is to alternatively switch the asymmetry between the views. In the case of asymmetric compression the solution is to alternatively switch the QPs of two views so that left/right eye dominant viewers do not notice a significant difference in the 3D video quality [21-22].

In this study the effect of eye dominance on the perceptual quality of stereo views with one synthesized view was studied for the case where the original high quality depth map is available. In practice, however, the depth map needs to be coded and transmitted and thus its quality is degraded due to compression. In the future, we plan to investigate this case too. Moreover, we will use a wide range of QP values to generate asymmetric compressed videos so that the effect of eye dominance is studied for different compression levels.

## 4. CONCLUSION

This study investigates the effect of eye dominance on the perceived quality of stereoscopic videos. A database of stereoscopic sequences containing representative types of asymmetry is prepared and the quality of these videos is subjectively evaluated by two groups of left and right-eye dominant subjects. Performance evaluations showed that eye dominance changes the mean opinion quality score by up to 16 % in the case of slight asymmetric video compression. The statistical difference is less for the other representative types of asymmetry.

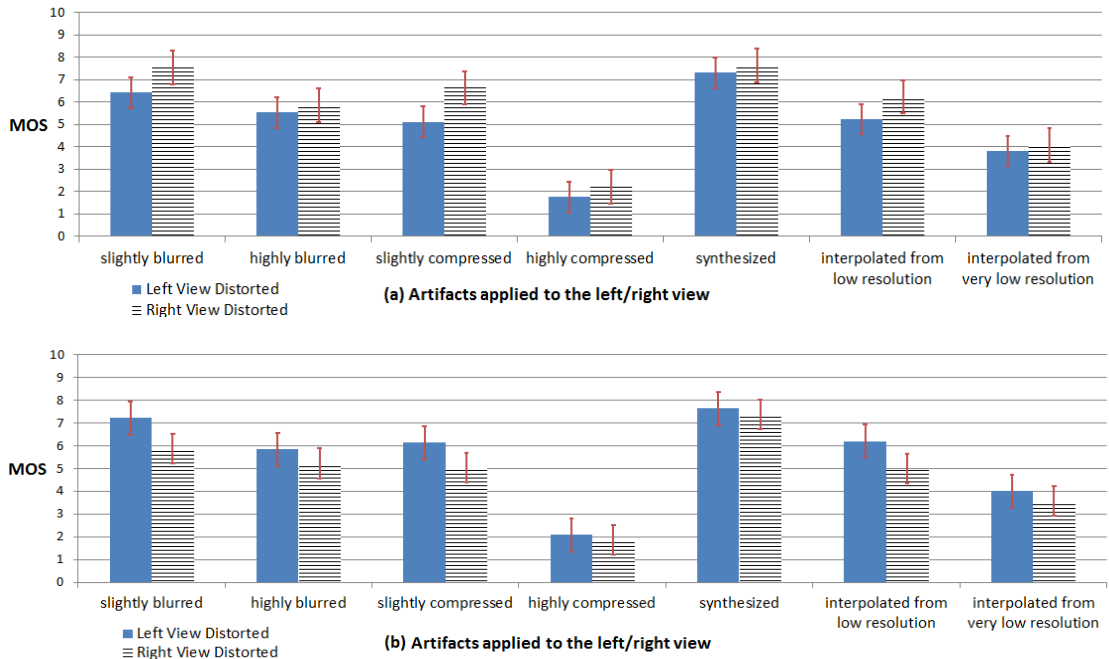

Fig. 2. Video quality scores reported by left eye dominant subjects (a) and right eye dominant ones (b).